\documentclass[apj]{emulateapj}
\usepackage[]{natbib}
\usepackage{graphicx}
\usepackage{color}

\shorttitle{WOCS LII: Old Open Cluster NGC6819}
\shortauthors{Yang et al.}

\begin{document}

\title{WIYN Open Cluster Study LII: Wide-Field CCD Photometry of the Old Open Cluster NGC 6819}

\author{Soung-Chul Yang\altaffilmark{1,4}}
\author{Ata Sarajedini\altaffilmark{2}}
\author{Constantine P. Deliyannis\altaffilmark{3}}
\author{Angela R. Sarrazine\altaffilmark{3}}
\author{Sang Chul Kim\altaffilmark{1}}
\author{Jaemann Kyeong\altaffilmark{1}}

\affil{$^1$Korea Astronomy and Space Science Institute, Daejeon 305-348, Republic of Korea}
\affil{$^2$Department of Astronomy, University of Florida, P. O. Box 112055,
    Gainesville, FL 32611}
\affil{$^3$Department of Astronomy, 319 Swain Hall West, 727 E. 3rd. St.,
Indiana University, Bloomington, IN 47401-7501}
\altaffiltext{4}{{\tt sczoo@kasi.re.kr}}

\begin{abstract}
We present a comprehensive photometric study of the old open cluster, NGC 6819 using 1$^\circ$$\times$1$^\circ$ field $VI$ MOSAIC CCD imaging taken with the WIYN 0.9m telescope. The resultant color-magnitude diagram (CMD) shows a well developed main sequence (MS) extending from $V$$\sim$14.5 mag down to our photometric limit of $V$$\sim$21 mag. Fitting theoretical isochrones with adopted values of the reddening and metallicity ($E(B-V)=$0.14, $[Fe/H]=+$0.09 dex) to the observed CMD yields a distance modulus of $(m-M)_{0}=$11.93$\pm$0.10 and an age of $\sim$ 2.6 Gyr for NGC 6819. Our wide-field imaging reveals that NGC 6819 is larger in areal extent ($R=$13$'$) than previously thought. The wide-field also benefits our estimate of the degree of field star contamination, and ultimately yields improved measurements of the structural parameters ($r_c=$2.80$'$, $r_t=$38.2$'$, and $r_h=$7$'$) and tidal mass of the cluster ($M_{tid}=$3542.4 $M_\odot$). The flattened luminosity and mass functions indicate that NGC 6819 has experienced mass segregation as a result of its dynamical evolution. Our variability study of the cluster blue straggler star (BSS) population using the Welch-Stetson variability index ($I_{WS}$) has revealed a number of variable BSS candidates.
\end{abstract}







\section{Introduction}
According to the latest compilation of known Galactic open clusters (i.e., Open clusters and Galactic structure, Version 3.2, http://www.astro.iag.usp.br/$\sim$wilton; Dias et al. 2012), now numbering over 2000, old clusters with an age of $>$ 1 Gyr are about 15\% of the sample accounting for a small fraction of the total open cluster population. This could be explained by the process of disruptive encounters with massive molecular clouds in the Galactic disk, which can easily destroy a typical open cluster (Spitzer 1958). Assuming a uniform destruction rate, the current census of old clusters implies that there once was a significant number of very old open clusters in the Galactic disk. Combined with the Galactic positions, their ages and metallicities provide a direct panoramic view of the early stages of formation and chemical evolution of the Galactic disk. These old clusters are also the key objects which can provide a possible link between old disk populations and the ancient Galactic globular clusters (Friel 1995). 

Given their practical importance in the study of Galaxy formation and evolution, several notable open cluster survey programs (i.e., WIYN Open Cluster Study (WOCS), Mathieu 2000; CFHT Open Star Cluster Survey, Kalirai et al. 2001a; Southern Open Cluster Study (SOCS), Kinemuchi et al. 2010; and Bologna Open Clusters Chemical Evolution (BOCCE) project, Bragaglia \& Tosi 2006; etc.) have been initiated and are actively being carried out for systematic investigations of the physical properties of the Galactic open clusters using homogeneous data sets. 

As a part of our continuing efforts in the WOCS series of studies, here we present a comprehensive photometric study of the old open cluster, NGC 6819 using 1$^\circ$$\times$1$^\circ$ field $VI$ MOSAIC CCD imaging. NGC 6819 is a relatively well studied open cluster (age $\sim$ 2.5 Gyr; Rosvick \& Vandenberg 1998, hereafter RV98; Kalirai et al. 2001b, hereafter K01, and references therein), but in recent years, it has gained renewed prominence because this cluster is one of the target fields of  NASA's $Kepler$ satellite mission. A couple of analyses of astroseismic data obtained from $Kepler$ observations of red giant branch stars (RGBs) in NGC 6819 have already been published (Stello et al. 2010; Basu et al. 2011). These studies provide model-independent estimates of the basic physical parameters (i.e., mass, radius, and log $g$) of RGB stars in NGC 6819. 

The 1$^\circ$$\times$1$^\circ$  CCD imaging presented herein is unique because it provides the largest sky coverage yet of NGC 6819 and its environs. 
This paper is organized as follows. Section 2 provides a short description of the observations and data reduction; in Section 3, we present the results of our analyses: Section 3.1 provides a detailed explanation of the cluster centering algorithm we have employed; Section 3.2 describes general trends in the cluster and field color-magnitude diagrams (CMDs); Section 3.3 explains our adoption of the reddening and metallicity values for the cluster, while Section 3.4 illustrates the measurements of distance and age; Section 3.5 provides a full description of the dynamical evolution of NGC 6819 using its surface density profile, luminosity and mass functions; Section 3.6 describes our attempt to detect variable blue stragglers; lastly Section 4 presents a summary of our results.
    
\vskip 1cm

\section{Observation and Data Reduction}

The observations of NGC 6819 were obtained at the WIYN 0.9m telescope at Kitt Peak National Obsevatory
on the night of 2000 July 9. Table 1 shows a log of the observations. The NOAO MOSAIC imager (8176$\times$8192 pixel$^{2}$) was used at the f/7.5 secondary to obtain a field of view of approximately
1$^\circ$$\times$1$^\circ$ with a plate scale of 0.4 arcsec pixel$^{-1}$. The typical
gain setting was 2.9 e$^-$ ADU$^{-1}$ with an average readout noise of 5.7 e$^-$. The images
were centered on the cluster and dithered between exposures to minimize errors due to 
flat-fielding and cosmetic defects as well as to fill in the gaps between the eight CCDs
that comprise the MOSAIC imager. 

The observed cluster images (15 $V$ + 15 $I$) were reduced following the standard procedure using the DAOPHOT II photometry package (Stetson \& Harris 1988). Ignoring the bad pixels, saturated stars, and defects, we selected about $\sim$ 1000 fairly bright, isolated stars in order to construct a high signal-to-noise point spread function (PSF) for each image. A PSF was constructed so that its shape varies quadratically with position across the field of view. Then, the instrumental magnitudes and positions of star-like objects were derived by fitting this PSF to all detected profiles using the ALLSTAR task. 

\begin{table}
\begin{center}
\caption{Observation Log. \label{tbl-1}}
\end{center}
\begin{tabular}{ccccccc}
\tableline\tableline
 &               &          & &  Exposure        &           & \\
 & Obs Date      &  Filter  & &   (sec)          &  Air Mass & \\
\tableline  
 & 2000 July 9   &   $V$    & &  7$\times$50     &   1.05    & \\
 &               &   $V$    & &  8$\times$200    &   1.05    & \\ 
 &               &   $I$    & &  7$\times$50     &   1.05    & \\
 &               &   $I$    & &  8$\times$200    &   1.05    & \\
\tableline
\end{tabular}
\end{table}

The next step included the calculation of aperture corrections for each image. To do so, $\sim$ 1000 bright stars were selected from each frame. All stars except these selected aperture correction stars were subtracted from the frame, at which point, photometry was performed on each aperture correction star using a large aperture size. The aperture correction was calculated by measuring the difference between the PSF magnitude and the total magnitude. We checked the degree of variation of the aperture correction with position on the image. There were no clear patterns or systematic variations in the aperture correction values with position on the images. Therefore we did not make any additional corrections for this effect. 

The calibration of the instrumental magnitudes was performed using standardized photometry of NGC 6819 provided by Sarrazine et al. (2003). The equations used for the transformation are the following : 

\begin{equation}
  V = v - 0.034 (\pm0.0006) \times (v-i) + 0.679 (\pm0.0009)  
\end{equation} 
\begin{equation}
  I = i - 0.008 (\pm0.0007) \times (v-i) + 0.320 (\pm0.0009)
\end{equation}

In this formulation, the instrumental magnitudes corrected for exposure time are shown as the lower case symbols, while the uppercase letters represent the standard system. 
The $rms$ errors of the fits in the $V$ and $I$ passband equations were 0.016 mag and 0.018 mag, respectively.

\begin{figure}
\epsscale{1.0}
\plotone{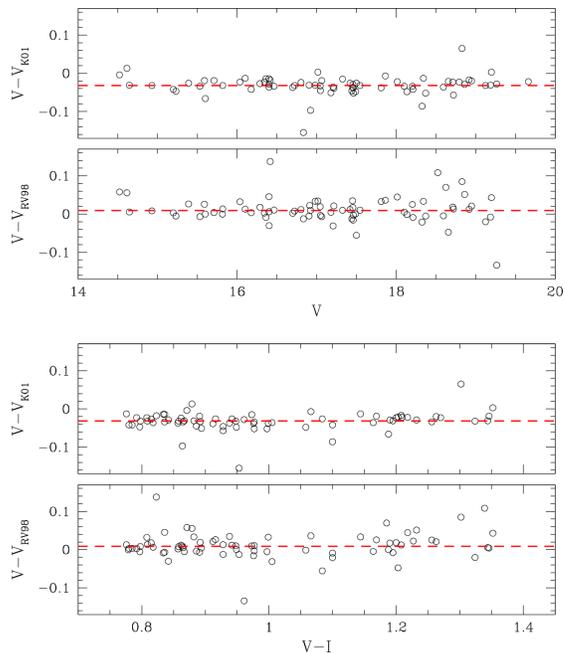}
\caption{The comparison of our photometry for NGC 6819 with previous studies of Rosvick \& Vandenberg 1998 (RV98) and Kalirai et al. 2001 (K01). This comparison includes only stars cross-identified in all the three studies and shows generally good agreement among them (see text for details).
 \label{fig 1}}
\end{figure}

We compared our photometry with literature data from RV98 and K01. For the stars cross-identified in all three studies (RV98, K01, and this study), the $V$ magnitude differences (this study - literature) were calculated. The results are presented in Fig. 1. In particular, the two upper panels show the magnitude differences as a function of $V$ magnitude from the present study, while the lower panels exhibit the magnitude differences as a function of $V-I$ color from this study. The comparison in Fig. 1 indicates that our photometry shows good agreement with those of RV98 and K01. The systematic differences in $V$ magnitude are of the order of 0.015 mag to RV98, and -0.035 mag to K01.

\section{Results}
\subsection{Cluster Centering}
We begin our analysis of NGC 6819 by determining the position of the cluster center. Locating an accurate cluster center is an important initial step for deriving robust physical parameters of star clusters. For instance, faulty centering would artificially flatten the inner part of the surface density profile of a star cluster, and this can cause significant systematic errors in any structural parameters such as core/tidal radii and total mass, and as a result important information on the dynamical evolution of the cluster could be compromised. 

The centering of Galactic open clusters is generally less straightforward than Galactic globular clusters because the former tend to be more asymmetric than the latter. In order to avoid significant biases in the centering, we employed three independent centering algorithms, and took the average of these results to derive an accurate center for the cluster. 

First, we used well-photometered stars which passed our stellarity criteria using the image-shape statistics provided by DAOPHOT II (i.e., 0.2 $<$ SHARP $<$ 0.5, -0.2 $<$ ROUND $<$ 0.2, $\chi$$^2$ $<$ 2.0). We counted these stars within narrow bands (80 pixels wide) in both the horizontal (X) and vertical (Y) directions across the images. These counts are shown in Fig. 2. We adopted the location at which the counts reached a maximum as the (X,Y) position of the cluster center.

\begin{figure}
\epsscale{1.0}
\plotone{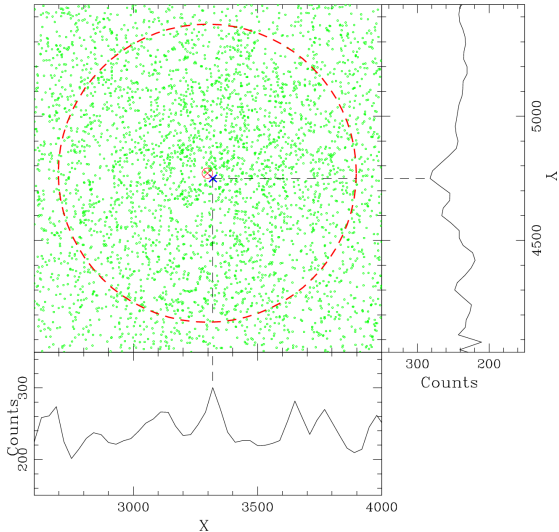}
\caption{Cluster centering via star counts within narrow bands (80 pixel wide) in the horizontal (X) and vertical (Y) directions across the image. The initial estimate of the cluster center is indicated as the blue cross, while our final estimate of the cluster center is marked by a red open-crossed symbol. The dashed circle represents the test region with a radius of 600 pixels used for our Monte Carlo centering method.  
 \label{fig 2}}
\end{figure}

The second centering algorithm we employed is called the ``four quadrants method'' (i.e., section 3 of K01). We adopted the (X,Y) location from the previous method as an initial estimate of the cluster center. Then we counted the number of stars in four equal quadrants around this center. We repeated the star counting by randomly moving the (X,Y) position of the center within a small circular area around the initial center until the number of stars in all four quadrants agreed. 

Finally, we implemented a Monte Carlo centering method similar to the algorithm employed by Mackey \& Gilmore (2003). We first placed a boxed area with a size (1200$\times$1200 pixels) large enough to cover the central region of the cluster. Next, we randomly selected an (X,Y) position as a test center. We calculated the surface density in a circular area with a radius of 600 pixels about this test center. We repeated this random selection of the test center location 10,000 times.  We consider the (X,Y) position that yields the highest surface density as the cluster center.  The results of these three independent centering algorithms agree reasonably well with each other. The average value of the (X,Y) positions provides our final estimate of the cluster center; we find (X,Y)=(3297.86, 4770.56) or (RA(J2000)=19:41:16.43, Decl(J2000)=+40:11:48.88). Our estimate of the centering errors in the X and Y directions is $\sim$ 25 pixels (10.6$''$) each. 

Our estimation of the cluster center shows excellent agreement with the recent calculation by Hole et al. (2009; RA=19:41:17.5, Decl=+40:11:47; hearafter H09) obtained using stars with moderate-to-high membership probability ($p > 50 \%$); however, it differs slightly from the center published by K01 (i.e., RA=19:41:17.7, Decl=+40:11:17) by $\sim$ 30$''$. Our tests have revealed that this difference is not significant enough to cause serious biases in the structural parameters of the cluster determined herein. Therefore we will adopt our value for the cluster center in the remainder of our analysis. 

\subsection{Color-Magnitude Diagram}
Figure 3 presents deep $VI$ color-magnitude diagrams (CMDs) in the region of NGC 6819. The two upper panels show CMDs for the cluster field defined by its visual extent ($R < 13'$) and the control field acquired from the outskirts ($27'<R<30'$) of the cluster with an area equal to that of the cluster field. These two CMDs provide useful insight to understand the degree of field star contamination (see section 3.5.1 for detail). In order to see the characteristic features of the cluster more clearly in the CMD, we extract stars from a central area ($R<6'$) centered around the cluster as shown in the lower left panel of Fig. 3; this shows a clean red giant branch (RGB) extending $\sim$1.5 mag brightward from a well-populated helium burning red clump (RC) at $V$$\sim$13 mag. There is a sparsely populated subgiant branch, a well-developed main sequence (MS) from $V$$\sim$14.5 mag down to our photometric limit of $V$$\sim$21 mag, and the signature of equal mass binary stars (0.75 mag brighter than the MS). We note, however, that  the observed equal-mass binary sequence in our $VI$ CMD is not as prominent as that of K01. The relative sparseness of our binary sequence combined with the degree of field star contamination made obtaining a robust estimate of the cluster binary fraction challenging. Therefore, we have decided to adopt the result of K01 (binary fraction $\sim$ 11\%) for the fraction of equal-mass binaries in NGC 6819.

\begin{figure}
\epsscale{1.0}
\plotone{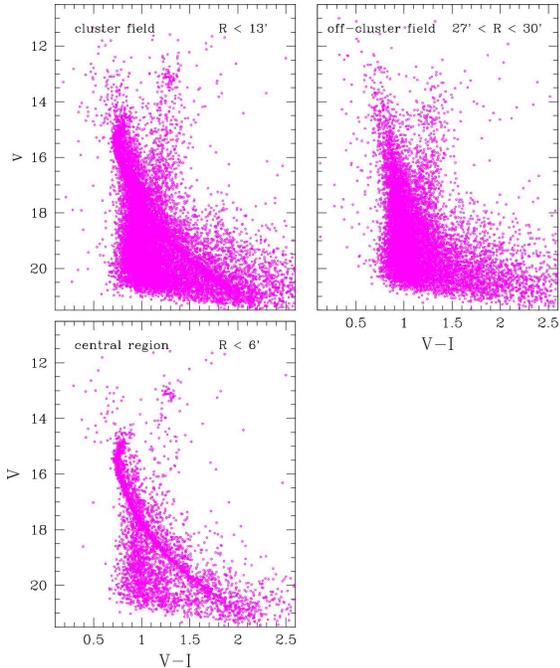}
\caption{$VI$ CMDs in the vicinity of NGC 6819 : ($upper$ $left$) cluster field CMD composed of stars within $R <$ 13$'$ from the cluster center, ($upper$ $right$) field star CMD obtained from an outer annulus (27.0$'$ $< R <$ 30$'$) with an area equal to that of the cluster field, ($lower$ $left$) central region CMD ($R <$ 6$'$) reveals the cluster principal sequences. 
\label{fig 3}}
\end{figure}

We also see the presence of a number of blue straggler stars (BSS), which are more luminous and bluer than the MS turnoff. BSSs are rejuvenated massive stars which likely form via stellar mergers either due to dynamical collisions of two MS stars or through the coalescence of close binaries, and mass transfer of an evolved star onto the less-massive MS partner in binary systems (Glebbeek et al. 2008). Some BSSs exhibit pulsation properties which obey period-luminosity and period-luminosity-color relations similar to those of classical Cepheids (Mateo 1993). The existence and pulsation properties of the BSS population in old Galactic open clusters might shed some light on their formation mechanism, and role in the dynamical evolution of the host cluster. We will discuss the membership and physical properties of the BSS population of NGC 6819 in section 3.6. 

\subsection{Reddening and Metallicity}
In order to derive the cluster age accurately by fitting theoretical isochrones to the observed CMD, we need well-determined cluster parameters such as reddening and metallicity. The early measurements of the interstellar reddening toward NGC 6819 in the 1970s spanned a substantial range between high ($E(B-V)$$\sim$0.30 : Lindoff 1972; Auner 1974) and low ($E(B-V)=$0.12 : Burkhead 1971) values. However as more observations of NGC 6819 have been accumulated, the reddening measurements from various independent studies have converged to the relatively low value of $E(B-V)$$\sim$0.15. The main reason for the high reddening measurements in the early studies is likely attributed to the photographic observations in the $U$ passband, which was known to be of low photometric quality. 

For the present study, we have adopted the reddening and metallicity values ($E(B-V)=$0.14$\pm$0.04;$[Fe/H]=$+0.09$\pm$0.03 dex) of NGC 6819 from the measurements of Bragaglia et al. (2001, hereafter B01). Using effective temperatures of three RC stars in NGC 6819 derived from line excitation, which is a reddening-free index, B01 measured the interstellar reddening toward the cluster. The reddening was calculated by comparing the observed $(B-V)$ colors with the intrinsic colors computed using the measured temperatures.  Any possible zero-point offset in the intrinsic colors derived by model atmospheres was also corrected using a sample of 12 bright RC stars with accurate $Hipparcos$ parallaxes. They selected this comparison sample to be similar to the stars of NGC 6819 in temperature and metallicity. B01's measurement of the interstellar reddening toward NGC 6819 agrees very well with other independent studies (Burkhead 1971; RV98; K01; Kang \& Ann 2002). In addition, their study appears to be the lone high-dispersion spectroscopic analysis of this cluster. It is for this reason that we consider the reddening and metallicity values of B01 to be the most robust measurements for NGC 6819 to date. 

\subsection{Distance and Age}
Applying the adopted values of the cluster reddening and metallicity in the previous section, we generated a set of isochrones using the Dartmouth Stellar Evolution Database (Dotter et al. 2007, 2008(D08)). The database program handles convection by following the general principles of the standard mixing length theory and a solar-calibrated mixing length, $\alpha_{ML}$=1.938. Convective core overshooting is treated in the same fashion as the $Y^2$ models developed by Demarque et al. (2004); the amount of convective core overshooting is parameterized as a multiple of the pressure scale height ($\lambda_P$), and is a function of stellar mass and composition. The database program assigns 0.05$\lambda_P$ of overshooting for the minimum stellar mass of 0.1$M_\odot$; then the amount of overshooting is linearly ramped from this value (0.05$\lambda_P$) with an increment of 0.1$\lambda_P$ as the stellar mass increases by 0.1$M_\odot$; models with masses greater than or equal to the minimum plus 0.2$M_\odot$ receive 0.2 $\lambda_P$ of overshooting.

The database uses a grid of PHOENIX model atmospheres (Hauschildt et al. 1999a, 1999b) for the range of $T_{eff}=$2,000 to 10,000 $K$ and log $g = -$0.5 to 5.5 to derive the surface boundary condition (BC). The computation of PHOENIX model atmosphere grids covers the wide range of metallicity and $\alpha$-element ratio ($-2.5 < [Fe/H] < +0.5$;  $-0.2 < [\alpha/Fe] < +0.8$). The effects of Helium enhancement to the physical structure and the stellar spectrum are also considered in the model calculation. We adopted the solar values of the $\alpha$-element ratio and He content ($[\alpha/Fe]=$0.0; $Y=$0.27) for the calculation of the isochrones.

\begin{figure}
\epsscale{1.0}
\plotone{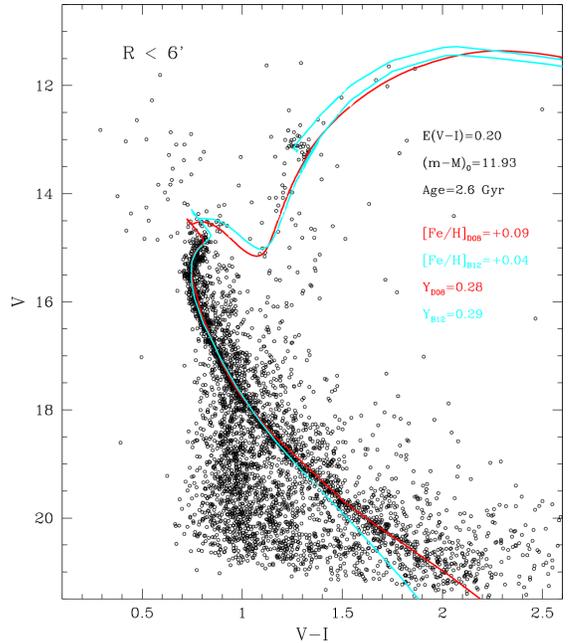}
\caption{$VI$ CMD of NGC 6819 along with the best-fitting theoretical isochrones derived from the Dartmouth Stellar Evolution Database (red, Dotter et al. 2007, 2008(D08)), and the PAdova \& TRieste Stellar Evolution Code (PARSEC; cyan, Bressen et al. 2012(B12)). The $E(V-I)$ value was calculated using the relation $E(V-I) =1.43E(B-V)$ from Sarajedini et al. (1999).
 \label{fig 4}}
\end{figure}

A series of theoretical isochrones were calculated for a fixed metallicity of $[Fe/H]=$ +0.09 dex, and ages ranging from 1 Gyr to 4 Gyr. The interpolation routine in the database produces isochrones with a fine age spacing of 0.1 Gyr. Doing so allows for more precise age constraints. These theoretical isochrones were transformed into the Johnson-Cousins photometric system using the filter transmission curves defined by Bessell (1990). Adopting the reddening, $E(B-V)=$0.14, and the initial estimate of distance modulus, $(m-M)_V=$12.30 taken from K01, we shifted the theoretical isochrones in the absolute magnitude-intrinsic color plane into the observed color-magnitude plane. Then, we varied the age and distance until a satisfactory fit was achieved between the observed CMD and the isochrones. 
Figure 4 shows our isochrone fits to the observed data of NGC 6819. We see that the 2.6 Gyr Darthmouth isochrone (red) with a refined distance modulus, $(m-M)_{V} =$12.36 yields the best description of the observed data. For comparison we also plot a best-fitting Padova isochrone (cyan) calculated from their latest stellar evolution code, PARSEC (Bressan et al. 2012; B12) for the given reddening, distance and age of the cluster. The Padova isochrone agrees well with the main features of the observed cluster CMD, such as the RGB, red clump (RC), and MSTO. However it starts to show offsets in the lower main-sequence ($V$ $>$ 18.5 mag) in the sense that the isochrone becomes systematically bluer than the observed sequence. The two model sequences also show a marginal offset at the subgiant branch, and upper-MS hook which is a characteristic feature among the old Galactic open clusters partly due to the effects of convective core overshooting. This offset is likely caused by the different treatment of convective core overshooting in each of the stellar evolution codes.  

The given amount of error ($\pm$0.04 mag) in the adopted reddening value yields a change in the derived distance modulus of $\pm$0.1 mag and a change in the derived age of $\pm$0.6 Gyr for a fixed $[Fe/H]$ value of 0.09 dex. A change in $[Fe/H]$ of $\pm$0.1 dex results in a change in the derived age of $\pm$0.3 Gyr and distance modulus of $\pm$0.08 mag for the adopted $E(B-V)$ value of 0.14 mag. This best-fitting Darthmouth isochrone seems to reproduce most of the characteristic features of the cluster CMD from the tip of the RGB down to the faintest low mass MS stars.


Our estimates for the age and distance modulus of NGC 6819 agree well with those determined from the recent analysis of Basu et al. (2011). Just like in this study, Basu et al. (2011) adopted the same reddening and metallicity values from Bragaglia et al. (2001). They measured a cluster age of 2.4 Gyr and a distance modulus of $(m-M)_V=$12.32 by employing a model-independent analysis of the astro-seismic quantity, $\Delta \nu$ using data obtained by NASA's $Kepler$ mission. Our result is also consistent with previous estimates in the literature ($(m-M)_V=$12.35, RV98; 12.30, K01; 12.11, Kang \& Ann 2002; 12.30, H09).

\begin{figure}
\epsscale{1.0}
\plotone{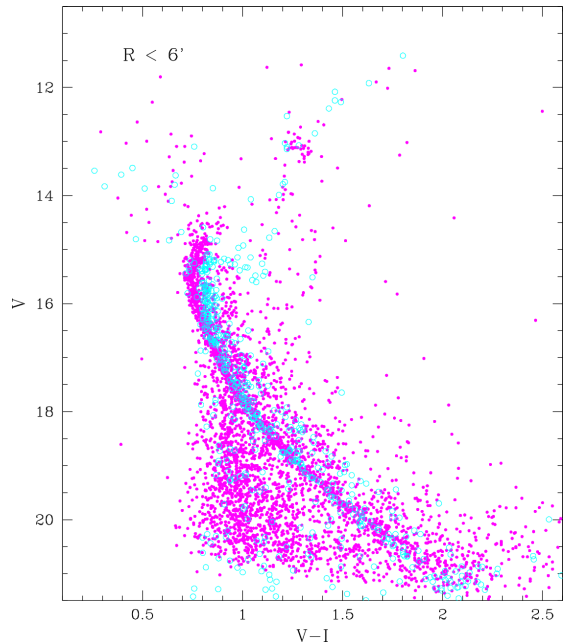}
\caption{A direct CMD comparison between NGC 6819 (solid circle) and M67 (open circle; age $\sim$ 4 Gyr) constructed using our estimate of the distance modulus of $(m-M)_{v}=$12.36, and the adopted interstellar reddening ($E(B-V)=$0.14; i.e., 0.20 in $E(V-I)$) from B01. The principal sequences of two clusters show very good agreement.  
 \label{fig 5}}
\end{figure}

The CMD in Fig. 5 shows a direct comparison of our photometry for NGC 6819 and that of M67  (Montgomery et al. 1993), which is one of the most studied old open clusters. This provides an alternative test of our adopted reddening and distance as well as the derived age for NGC 6819. In order to perform this direct CMD comparison, we need to adopt a reddening and distance for M67, for which we rely on the results of Sarajedini et al. (1999), who found $E(B-V)=$0.04, $(m-M)_V=$9.69.
These values along with the reddening and distance of NGC 6819 ($E(B-V)=$0.14, $(m-M)_V=$12.36) allow us to produce the CMD comparison shown in Fig. 5. We see that there is a nearly exact match of the unevolved main sequences of these two clusters, and a reasonable separation of the main-sequence turnoffs predicted by the age difference ($\Delta t_{age}$$\sim$2 Gyr) between them. The red giant branches are also very well aligned indicating that the two clusters have similar metallicities within roughly 0.1-0.2 dex. Thus this CMD comparison strengthens the validness of our estimates of the distance modulus and age for NGC 6819. 


\subsection{Dynamical Evolution of NGC 6819}
Now we direct our attention to the dynamical properties of the cluster. 
These have previously been investigated by K01 and Kang \& Ann (2002). K01 conducted a wide field (42$'$$\times$28$'$) $BV$ imaging survey of NGC 6819 using the 3.6m Canada-France-Hawaii Telescope (CFHT) equipped with the CFH12K mosaic CCD. The structural parameters such as the core and tidal radii ($r_c=$1.75 pc; $r_t=$17 pc) of the cluster were derived through a King model fit to the observed density profile constructed with star counts. They found an almost flat present-day luminosity function, and the clear signature of mass segregation from the analysis of radial CMDs and mass functions; their results indicate that NGC 6819 has dynamically evolved, and now contains mostly high-mass stars in the inner regions and lower mass stars at larger radii. Subsequently, Kang \& Ann (2002) presented $VI$ CCD photometry of the cluster. Their observational data covered an area of 18$'$$\times$18$'$ on the sky, which is smaller than K01's field but still covered the visual extent of the cluster. By analyzing the derived luminosity function and radial density profile, they confirmed K01's result that the cluster clearly exhibits mass segregation as a result of its dynamical evolution. 

The NGC 6819 photometry presented herein represents the widest field coverage of this cluster to date. As such, it covers virtually the entire dynamical extent of the cluster.
This provides an excellent opportunity to investigate the dynamical structure of the cluster in great detail. In the following two subsections, we describe our analysis of the surface density profile, luminosity/mass function, and mass segregation in the cluster.  

\subsubsection{Surface Density Profile and Cluster Mass}
In order to construct the surface density profile of the cluster, we used well-photometered stars that had passed our stellarity criteria of the image-shape statistics (i.e., 0.2 $<$ SHARP $<$ 0.5, --0.2 $<$ ROUND $<$ 0.2, $\chi^2$ $<$ 2.0; see section 3.1), and fall in the magnitude range, 15 $< V <$ 20 (i.e. the corresponding mass range is 0.73 -- 1.52 $M_{\odot}$).
A series of concentric circles around the cluster center (RA(J2000)=19:41:16.43, Decl(J2000)=+40:11:48.88, see section 3.1) was used to construct the surface density profile shown
in the upper panel of Fig. 6. The stellar density appears to flatten out at $\sim$13$'$ and becomes indistinguishable from the background field beyond this radius. We consider this region to constitute the visual extent of the cluster, ``$r_{cl}$'', which is the largest yet identified for this cluster (e.g., K01 measured a cluster extent of $\sim$9.5$'$). The surface density profile of the cluster shown in the lower panel of Fig. 6 includes corrections for photometric incompleteness (see the next section for detail) and for the density of the background field. The average background density was calculated in the outskirts ($25' < R < 30'$) of the cluster.

\begin{figure}
\epsscale{0.9}
\plotone{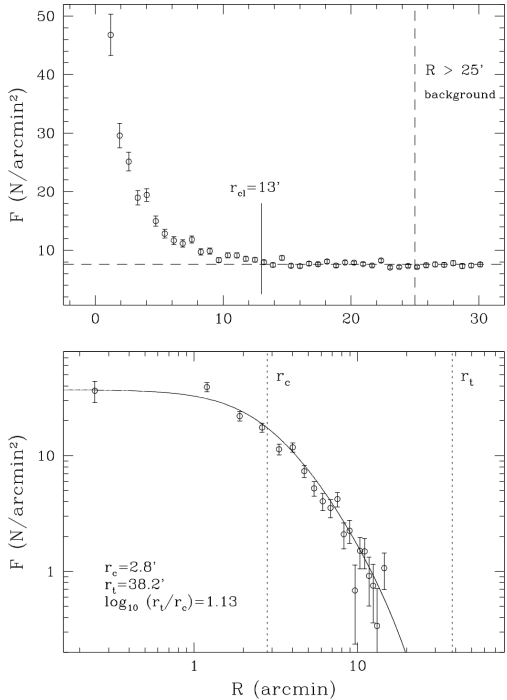}
\caption{Completeness corrected radial density profile ($upper$ panel) of NGC 6819 reveals the apparent extent of the cluster ending at $R$ $\sim$ 13$'$. The best-fit King model ($lower$ panel) very well describes the field star subtracted surface density profile. Error bars represent 1$\sigma$ Poission errors in the star counts.    
 \label{fig 6}}
\end{figure}

Once the background subtracted photometrically complete surface density profile was constructed, we fitted the following empirical King model (1962) to this profile. 

\begin{equation}
  f(r) = k \Bigg[\frac{1}{[1+(r/r_c)^2]^{1/2}}  - \frac{1}{[1+(r_t/r_c)^2]^{1/2}} \Bigg]^2,
\end{equation}
where $f(r)$ is the number density as a function of $r$, $r_c$ is the core radius, $r_t$ is tidal radius, and $k$ is a constant related to the central surface density. We employed a Maximum likelihood method powered by a genetic algorithm PIKAIA (Charbonneau 1995) to derive the best-fit structural parameters of NGC 6819. We obtained $r_c=$ 2.80$'$ ($\sim$ 2 $\pm$ 0.10 pc), $r_t=$38.2$'$ ($\sim$ 27 $\pm$ 1.2 pc), and log $(r_t/r_c)=$1.13. The quoted errors in $r_c$ and $r_t$ reflect the amount of errors propagated from the expected error of 0.10 mag in our distance modulus measurement.  

Next we can calculate the mass of the cluster within its tidal radius. 
Accurate measurements of the current masses of star clusters are crucial for understanding cluster survival and destruction processes in the Galactic disk. According to the classical theory of  Galactic dynamics, the tidal force exerted on star clusters reaches its maximum when the clusters are passing through their perigalactic positions. The tidal radius of a cluster is defined as the inner Lagrangian point where the Galactic gravitational potential exactly equals the gravitational potential of the cluster. Using this definition of the tidal radius of a star cluster and the point-mass Galactic model, King (1962) derived a relation between the tidal radius and the mass of a star cluster as follows,

\begin{equation}
  r_t = R_p \Big[\frac{M_c}{(3+e)M_g}\Big]^{1/3}, 
\end{equation}
where $R_p$ is the perigalactic distance of the cluster from the center of the Milky Way, $e$ is the eccentricity of the cluster orbit, and $M_g$ is the mass of the Milky Way within $R_p$. This simple version of the formula has been used in the calculation of total cluster mass using the tidal radius obtained from fitting a King model to the observed surface density profile, or alternatively, the tidal radius by assuming the total mass of the cluster. However, both theoretical considerations and numerical computations (Innanen et al. 1983; Keenan 1981) call for caution with regard to the use of this equation. This is because the tidal radius in this formula is not the same as the tidal radius derived from King model fitting to the observed surface density profile. The tidal radius calculated from the point-mass Galactic model (Equation (4)) refers to the semi-minor axis of the elongated tidal surface of the cluster, while the tidal radius obtained from the King model fitting represents the radius of the isothermal sphere at which the cluster stellar density vanishes into the Galactic field. This difference means that the semi-minor axis of the elongated tidal surface is 2/3 of the radius ($r_{t,King}$) of the isothermal sphere (i.e. the distance from the cluster center to the inner Lagrangian point). Therefore, the corrected tidal radius formula should be,

\begin{equation}
 r_{t,King} = \frac{3}{2} R_p \Big[\frac{M_c}{(3+e)M_g}\Big]^{1/3}
\end{equation}
By re-arranging the above equation, we obtain

\begin{equation}
 M_c = (3+e)M_g\Big[\frac{2}{3} \frac{r_{t,King}}{R_p}\Big]^3
\end{equation}

In order to calculate the tidal mass of NGC 6819, we adopted $R_p=$7.6 kpc as its perigalactic distance and $e=$0.03 as the orbital eccentricity of the cluster from Wu et al. (2009). The Galactic mass $M_g$ within $R_p$ can be calculated using $M_g = V_0^2 R_p / G = 8.8\times10^{10} M_\odot$ where $V_0=$220 km s$^-1$ is the circular rotation speed of the Milky Way. 
Using these inputs into Equation (6), we obtained the tidal mass of NGC 6819 to be $M_{c,tid} =$ 3542.4$^{+536.8}_{-451.6}$ $M_{\odot}$. We also estimated the total mass of the cluster by integrating the global mass function which is derived from the stellar mass-luminosity relation of the best fitting stellar evolution model (see section 3.5.3). Since the mass function was obtained from the cluster luminosity function with the help of theoretical models of stellar evolution, the total mass derived by this method is referred to as a ``photometric mass'', $M_{phot}$. We obtained $M_{c,phot} =$ 2010 $M_{\odot}$ which agrees fiarly well with K01’s calculation ($\sim$2600 $M_{\odot}$) of the photometric mass of NGC 6819. The estimates of the total mass of NGC 6819 from these three values yield the mean mass of NGC 6819 of $< M_c >$ $\sim$ 2700 $M_{\odot}$. 



\subsubsection{Luminosity Function} 
We constructed the present-day luminosity function (LF) using the cluster MS stars with $V$ magnitudes in the range $15 < V < 20$. The number of cluster MS stars was statistically estimated by the following method. We defined the MS envelopes in the $VI$ CMD obtained from the central region ($R < 6'$) of the cluster with two loci aligned with the blue and red edges of the MS. Using this MS envelope, we drew stars from both the cluster region ($R < 13'$) and an off-cluster region ($25' < R < 30'$) with an area equal to that of the cluster field.

\begin{figure}
\epsscale{0.8}
\plotone{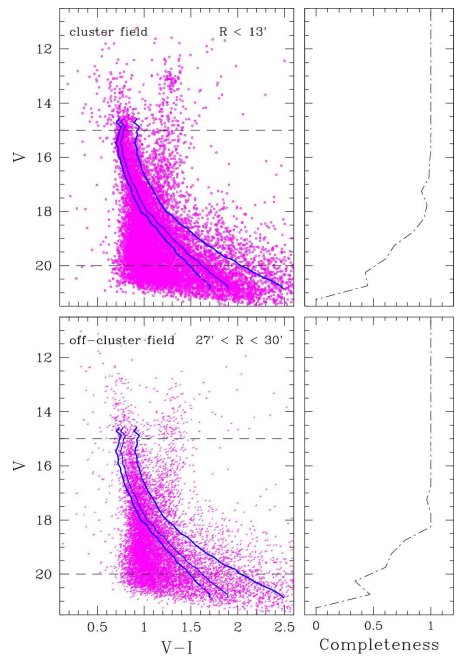}
\caption{Our selection of main sequence stars ($left$ panels) within the two slanted lines for the construction of the luminosity and mass functions of NGC 6819. The line between the envelopes illustrates the main sequence fiducial of the cluster derived from the central region CMD ($R <$ 6$'$). The photometric completeness profiles ($right$ panels) of both cluster and off-cluster fields show a similar pattern indicating that the crowding does not negatively affect the photometric completeness. 
 \label{fig 7}}
\end{figure}

We need to correct for photometric incompleteness in order to properly interpret the cluster LF. This was done through a series of artificial star tests. We employed the ADDSTAR routine in DAOPHOT II to generate $\sim$2,000 artificial stars with known magnitude and position. These were evenly distributed on each $V$ and $I$ program image. These test images were reduced using the same procedure as the original images. Our artificial star tests revealed that both the cluster and off-cluster fields show very similar photometric completeness profiles, and have a very similar overall recovery rate of $\sim$80 \%. The selection of the cluster MS stars and the photometric completeness as a function of $V$ magnitude are illustrated in Fig. 7. The photometric completeness correction factors are summarized in Table 2.

\begin{table}
\begin{center}
\caption{Photometric completeness for the cluster and background fields. \label{tbl-2}}
\end{center}
\begin{center}
\begin{tabular}{ccccccc}
\tableline\tableline
         & Range        &        &    Correction Factor                    &  \\
         &              &        &   $1/f=N_{in}/N_{out}$   &              &  \\
         & $V$(mag)     &        &   $1/f_{cl}$             &  $1/f_{bg}$  &  \\
\tableline
         & 14.0-14.5    &        &   1.00     &    1.00     &  \\
         & 14.5-15.0    &        &   1.00     &    1.00     &  \\      
         & 15.0-15.5    &        &   1.00     &    1.00     &  \\ 
         & 15.5-16.0    &        &   1.00     &    1.00     &  \\
         & 16.0-16.5    &        &   1.01     &    1.00     &  \\      
         & 16.5-17.0    &        &   1.05     &    1.00     &  \\      
         & 17.0-17.5    &        &   1.09     &    1.04     &  \\     
         & 17.5-18.0    &        &   1.03     &    1.00     &  \\      
         & 18.0-18.5    &        &   1.06     &    1.00     &  \\      
         & 18.5-19.0    &        &   1.18     &    1.29     &  \\     
         & 19.0-19.5    &        &   1.46     &    1.53     &  \\     
         & 19.5-20.0    &        &   1.65     &    1.65     &  \\     
         & 20.0-20.5    &        &   2.32     &    2.95     &  \\     
         & 20.5-21.0    &        &   2.22     &    2.13     &  \\    
\tableline
\end{tabular}
\end{center}
\end{table}

\begin{figure}
\epsscale{0.75}
\plotone{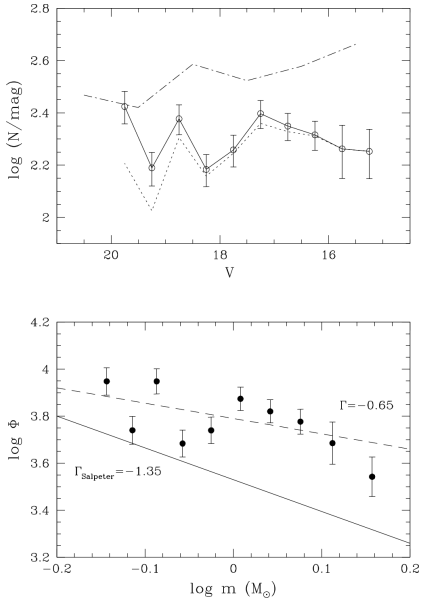}
\caption{Present-day luminosity function (LF) of NGC 6819 ($upper$ panel). Solid line illustrates the photometric incompleteness corrected LF, while the dotted line is the LF before the corrections. K01's LF (dot-dashed line; Fig.9 in their paper) is also plotted for comparison. Both LFs appear to be almost flat, though K01's LF shows a very weak upward slope. The lower panel describes the present-day mass function (MF) of NGC 6819 which was derived from the LF ($upper$ panel) using the stellar mass-luminosity relations defined in the Dartmouth isochrone database. The present-day MF of NGC 6819 exhibits a shallower slope as compared to the Salpeter IMF (solid line) indicating that this cluster has lost many of its low-mass stars via dynamical evolution. Error bars represent the total errors that include Poisson statistics and photometric incompleteness. The dashed line shows a weighted linear least-squares fit for the given mass range. 
 \label{fig 8}}
\end{figure}

The upper panel of Fig. 8 shows a comparison between our LF (solid line) and that of K01 (dot-dashed line). The zero-point offset between these two luminosity functions seems to be  caused by the different size of the main-sequence envelopes used for the selection of MS stars in each study. The error bars on each point reflect the uncertainties due to  Poisson statistics and the photometric incompleteness correction. We followed K01's recipes for the estimation of the errors in the global luminosity and mass functions.  The standard deviation in $N$, $\sigma_{N}$ is derived from the error propagation equation (Bevington \& Robinson 1992; equation 3.14, p43) as follows : 

\begin{equation}
\sigma_{N}^2=\frac{N_{obs}}{f^2}+\frac{(1+f)N_{obs}^2}{N_{in}f^3}.
\end{equation}

where $N$ is incompleteness corrected count defined by $N = N_{obs}/f$; $f$ is photometric incompleteness given by $f=N_{out}/N_{in}$ (i.e., Therefore the completeness correction factor should be the inverse of the incompleteness, $1/f$). 

These two LFs generally agree in the sense that both exhibit almost flat slopes, though K01's LF tends to decrease slightly toward faint magnitudes.
The slope of the present-day LF of a given star cluster can serve as an excellent diagnostic tool for examining the dynamical state of clusters. This flat or very weak slope of the present-day LF of NGC 6819 indicates that the cluster is old enough to have lost a significant portion of its low-mass stars through evaporation as it has dynamically evolved. 

\subsubsection{Mass Function and Mass Segregation}
Now we attempt to derive the present-day mass function (MF) of NGC 6819 to further investigate the dynamical evolution of the cluster. 
The MF is defined as the number of stars per unit logarithmic mass interval, $\phi$(log $m$)$=$$dN/d$log $m$ and often expressed by the power law, $\phi$(log $m$) $\propto$ $m^\Gamma$, where the slope of the MF is given as $\Gamma=d$log $\phi$(log $m$)/$d$log $m$. The traditional value of $\Gamma$ for the Salpeter IMF is $\Gamma=-$1.35. The present-day LF of NGC 6819, obtained in the previous section, was converted into the MF using the stellar mass-luminosity relation defined by the theoretical stellar evolutionary tracks of Dotter et al. (2007, 2008). The lower panel of Fig. 8 shows the present-day MF of the cluster MS stars drawn from the cluster region ($R < 13'$). The value of the cluster MF slope, $\Gamma=-$0.65 was computed using linear least-square fitting as shown as a dashed line. The Salpeter IMF ($\Gamma=-$1.35) is also plotted in the same plane for comparison. We see that the present-day MF of NGC 6819 is shallower than the Salpeter IMF. Again we interpret this flattened cluster MF as being mainly due to the evaporation of low-mass stars during the dynamical evolution of the cluster.

\begin{figure}
\epsscale{1.0}
\plotone{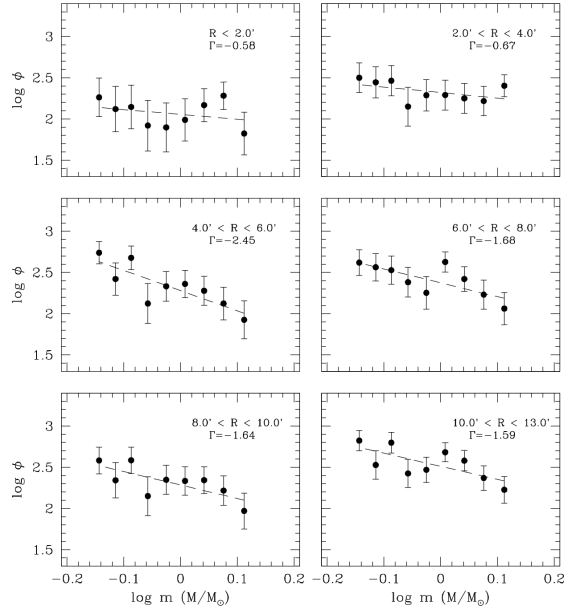}
\caption{Radial MFs for six annuli surrounding NGC 6819. 
The figure shows that the slopes ($\Gamma$) of the MFs appear to be flattened near the central regions but become steeper at larger radii until $R=$ 8$'$, and remain constant in the outskirts of the cluster. This provides support for the presence of mass segregation in NGC 6819.  
 \label{fig 9}}
\end{figure}

In Fig. 9, we plot radial MFs. The slope of the MFs tends to be flatter in the central regions and appears to be steeper at larger radii, indicative of the presence of mass segregation in the sense that high-mass stars preferentially lie near the cluster center. The resultant luminosity and mass functions with detailed statistics are summarized in Table 3.

\setlength{\tabcolsep}{3.0pt}
\begin{deluxetable}{rclcccc}
\tablecaption{The present-day luminosity and mass functions for NGC 6819 \label{tbl-3}}
\tablehead{
\colhead{Range} & \colhead{} & \colhead{} & \colhead{Count} & \colhead{} & \colhead{LF} & \colhead{MF} \\
\colhead{$V$(mag)} & \colhead{Mass($M_\odot$)} & \colhead{$<M>$} & 
\colhead{$N$} & \colhead{$\sigma_{N}$} & \colhead{log $N$} & \colhead{log $\phi$}  
}
\startdata
   15.0--15.5 &  1.5204--1.3509 &  1.4357 & 179.00 & 38.22 & 2.2529 & 3.5424     \\     
   15.5--16.0 &  1.3509--1.2384 &  1.2946 & 183.00 & 42.16 & 2.2625 & 3.6851     \\  
   16.0--16.5 &  1.2384--1.1434 &  1.1909 & 207.05 & 26.49 & 2.3161 & 3.7765     \\  
   16.5--17.0 &  1.1434--1.0578 &  1.1006 & 223.65 & 26.64 & 2.3496 & 3.8204     \\  
   17.0--17.5 &  1.0578--0.9795 &  1.0187 & 249.61 & 30.44 & 2.3973 & 3.8739     \\  
   17.5--18.0 &  0.9795--0.9078 &  0.9437 & 181.28 & 25.07 & 2.2583 & 3.7396     \\  
   18.0--18.5 &  0.9078--0.8441 &  0.8760 & 152.64 & 21.49 & 2.1837 & 3.6837     \\  
   18.5--19.0 &  0.8441--0.7934 &  0.8187 & 238.36 & 31.26 & 2.3772 & 3.9477     \\
   19.0--19.5 &  0.7934--0.7436 &  0.7685 & 154.76 & 22.72 & 2.1897 & 3.7398     \\  
   19.5--20.0 &  0.7436--0.6940 &  0.7188 & 265.65 & 37.72 & 2.4243 & 3.9479     \\
\enddata
\tablecomments{\footnotesize $N$ represents the number of cluster main sequence stars after correcting the photometric incompleteness and field star contamination.}
\end{deluxetable}

To further evaluate this assertion quantitatively we calculated the half-mass relaxation time for the cluster using the following equation derived by Spitzer (1987): 

\begin{equation}
 t_{rh}= \frac{0.183M^{1/2}r_h^{3/2}}{mG^{1/2}ln(0.4N)}
\end{equation}
where $M$ is the total mass of the cluster, $m$ is the mean stellar mass, $r_h$ is the radius containing half of the cluster mass, and $N$ is the number of cluster members. We determined the half-mass radius using the cumulative mass distribution of the cluster stars. Figure 10 indicates that 50\% of the cluster mass is contained within $\sim$ 7$'$ ($\sim$ 4.95 pc) which is very similar to half of the cluster extent. The total number of cluster stars was calculated by dividing the total mass of the cluster ($M=$ 2700 $M_\odot$, Sec. 3.5.1) by the mean stellar mass ($m$ $\sim$ 1 $M_\odot$ obtained from the global MF). By inserting these numbers into the equation above, we obtained the half-mass relaxation time of $t_{rh}$ $\sim$ 200 Myr. Compared to the age  of the cluster (2.6 Gyr), this relaxation time is about 10 times smaller, again indicating that NGC 6819 is old enough to be dynamically evolved and that the observed mass segregation is the result of its  dynamical evolution.

\begin{figure}
\epsscale{1.0}
\plotone{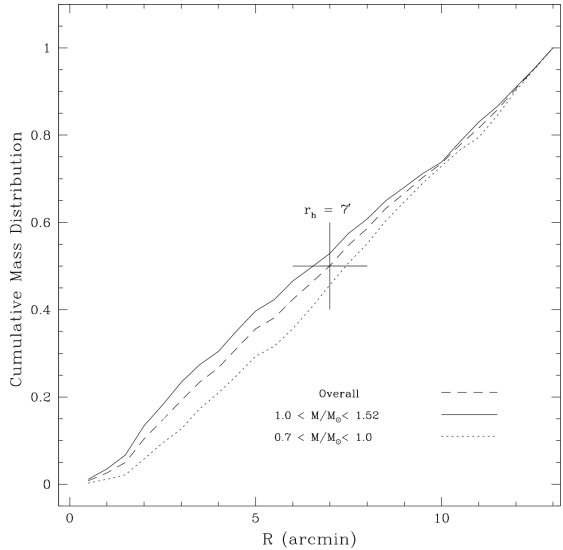}
\caption{Cumulative mass distribution of NGC 6819. This figure reveals that about 50\% of the cluster mass is confined within a radius of $r_h=$7$'$. 
 \label{fig 10}}
\end{figure}

\subsection{Variable Blue Stragglers}
Searching for pulsating variable BSSs in open clusters is an important subject in stellar astrophysics because the pulsational properties of BSSs can be used to derive their basic physical properties such as mass and radius; they also shed light on the formation mechanism of BSSs in the relatively sparse environments of open clusters. However our current knowledge of the periods and amplitudes of pulsating variable BSSs in open clusters is still highly uncertain.  
In his comprehensive review of BSSs, Mateo (1993) categorized variable BSS populations into two types : pulsating BSSs ($\delta$ Scuti or SX Phe type depending on its age) and eclipsing BSSs (mostly W UMa type). In particular, the pulsating BSSs are considered to be a class of Dwarf Cepheids (DCs) and appear to follow period-luminosity and period-luminosity-color relations similar to those of classical Cepheids. They pulsate with either fundamental or first-overtone mode periods ranging from $\sim$ 0.03 to 0.25 days, and exhibit very small amplitudes of 0.005 -- 0.8 mag. 

Previously, Kaluzny \& Shara (1988; hearafter KS88) found three variable star candidates which are likely short-period contact binaries in a 4$'$$\times$4$'$ field centered on NGC 6819. Among these three, there is one bright star (i.e. V2 in their list) included in the photometry of our study. 
Street et al. (2002, 2005) also detected a number of eclipsing binaries from a much wider cluster field ($\sim$ 11$'$$\times$22$'$). They confined their search to variable stars in a magnitude range fainter than the main sequence turnoff. 
Most recently, Talamentes et al. (2010; hereafter T10) presented a detailed variability study of this cluster. They reported the detection of three bright blue straggler variables. The brightest one appears to be a $\delta$ Scuti type pulsator, however its color is slightly bluer than the blue edge of the instability strip of $\delta$ Scuti variables. Star A494 (from their table) turned out to be the same star as V2 discovered by KS88, while the other one is likely a newly detected variable BSS probably in the process of merging. 

We carefully investigated the variability of our BSS candidates using the Welch-Stetson variability index ($I_{WS}$, Welch, \& Stetson 1993). We selected probable cluster BSS candidates from the $VI$ CMD using our color-magnitude criteria (11.5 $< V <$ 15.5; 0.2 $< V-I <$ 0.7). This selection of cluster BSS candidates was restricted to within the half-mass radius ($r_h=$7$'$) of the cluster in order to increase the chances of including true cluster members. Then, we calculated $I_{WS}$ values for these BSS candidates.

\begin{figure}
\epsscale{1.0}
\plotone{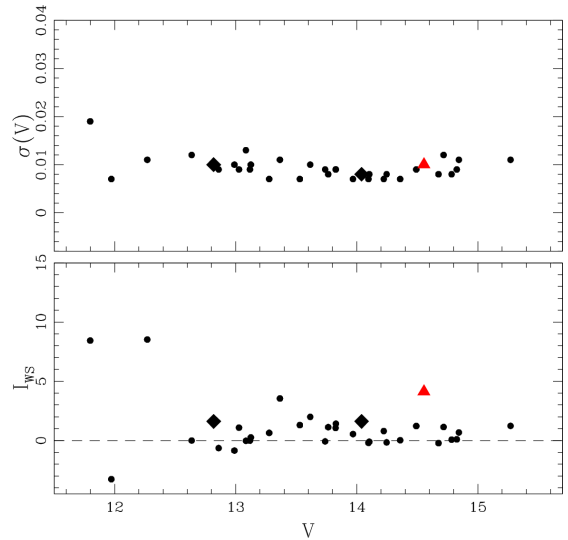}
\caption{Upper panel shows that the photometric errors of our BSS candidates are very small ($\sigma(V)$ $<$ 0.02). Lower panel presents the Welch-Stetson variability index ($I_{WS}$) as a function of $V$ magnitude for the BSS candidates. Two solid diamonds are the variable BSSs studied by T10. 
The probable $\delta$ Scuti type pulsating variable candidate detected in this study is marked by red triangle. This figure shows that many BSS candidates exhibit a moderate variablity (1.0 $<$ $I_{WS}$ $<$ 2.0). However, spuriously high $I_{WS}$ values for a couple of bright stars were likely produced due to the small number of observations. 
 \label{fig 11}}
\end{figure}

Figure 11 presents the $I_{WS}$ index as a function of $V$ magnitude. Filled circles represent the probable cluster BSSs, while the variable BSSs confirmed in the previous studies (KS88; T10) are marked by filled diamonds. 
A probable $\delta$ Scuti type variable candidate newly detected in this study is shown as a red triangle (see the following paragraphs). According to Fig. 11, the photometric errors are less than 0.02 mag (see $upper$ panel) for all of our BSS candidates. We also see that the majority of our BSS candidates appear to be non-variable stars ($I_{WS} <$ 1.0), while some show moderate variability (1.0 $< I_{WS} <$ 2.0). 
There are two bright stars with large values of $I_{WS}$. They may not be bona fide variables because spuriously large values of $I_{WS}$ could be the result of the small number of observations (i.e., $n_{obs} <$ 6 for these stars). 
Figure 12 is a zoomed-in $VI$ CMD showing these BSS candidates. We marked with crosses those BSSs whose $I_{WS}$ values are greater than 1.0. We adopted the pulsational instability strip from T10, originally derived by Pamyatnkh (2000), and overlaid it on the CMD in order to isolate the pulsationally variable BSS candidates. We see that many BSSs with moderate variability (i.e., $I_{WS}$ $>$ 1.0) are located within this pulsational instability strip indicating that some of these could possibly be $\delta$ Scuti type, short-period pulsators. 

\begin{figure}
\epsscale{1.0}
\plotone{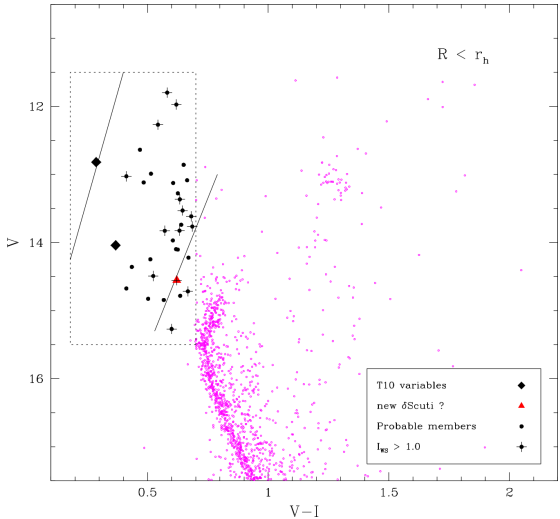}
\caption{Zoomed-in CMD for the BSS candidates in NGC 6819. We use the same symbols as in Fig. 11. The BSS candidates whose $I_{WS}$ value is greater than 1.0 are marked by crosses. The newly detected $\delta$ Scuti candidate is located on the red edge of the pulsational instability strip. The obtained period and amplitude values for this star seem to agree with those of typical $\delta$ Scuti type pulsating variables (0.03 $< P <$ 0.25 days; 0.005 $< Amp <$ 0.8 mag).  
 \label{fig 12}}
\end{figure}

\begin{figure}
\epsscale{0.7}
\plotone{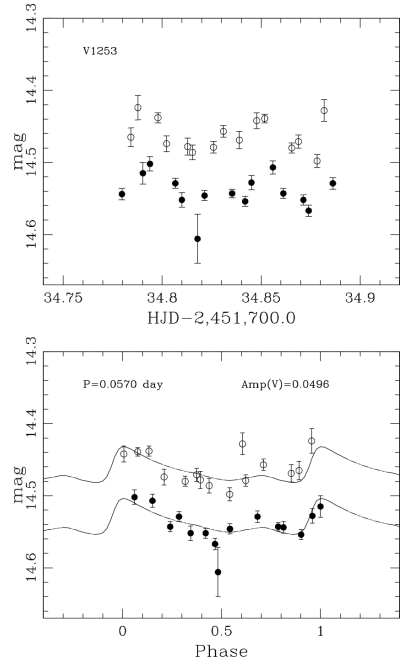}
\caption{Raw ($top$ panel) and best-fit light curve ($bottom$ panel) for the probable $\delta$ Scuti type variable candidate in NGC 6819. The best-fit light curve parameters, such as period, amplitude, maximum epoch, and mean magnitude, were derived by our template light curve fitting routine, RRFIT (see Yang \& Sarajedini 2012 for detail).
 \label{fig 13}}
\end{figure}

We attempt to uncover possible $\delta$ Scuti type pulsators by running our template light curve fitting routine, RRFIT (Yang \& Sarajedini 2012), on all of the BSSs with moderate variability ($I_{WS}$ $>$ 1.0) and 30 individual photometric measurements (15 $V$s + 15 $I$s). 
Figure 13 presents the best result from this exercise. We see what appears to be a probable $\delta$ Scuti type BSS detected by our template fitting analysis. According to its best-fit light curve, this star appears to be a fundamental mode pulsator with a short period and a small amplitude. The derived period and amplitude fall in the period-amplitude range for typical $\delta$ Scuti type pulsators (0.03 $< P <$ 0.25 days; 0.005 $< Amp <$ 0.8 mag) consistent with the pulsational properties of the short-period pulsating BSSs found in M67 (Gilland \& Brown 1992: $P$$\sim$0.06 days, $Amp(V)$$\sim$ 0.02 mag). However we must stress that our estimates of the period and amplitude should be treated with caution because the observational data in this study are not optimized for the study of variable stars. The list of our BSS candidates in NGC 6819 is summarized in Table 4.

\setlength{\tabcolsep}{3.5pt}
\begin{deluxetable*}{lrrccccl}
\tabletypesize{\footnotesize}
\tablecaption{The probable BSS candidates of NGC 6819 with a moderate variability ($I_{WS}$ $>$ 1.0)\label{tbl-4}}
\tablewidth{0pt}
\tablehead{
\colhead{ID}          &  \colhead{RA(J2000)}          &  \colhead{Decl(J2000)}      & 
\colhead{$V$}         &  \colhead{$V-I$}              &  \colhead{$I_{WS}$}         &
\colhead{$\sigma(V)$} &  \colhead{$Note$}  
}
\startdata
V234303  &  19:41:44.98  &  40:12:50.59 & 11.799 & 0.582 & 8.4335  & 0.019 &          \\
    V65  &  19:41:44.82  &  40:08:02.08 & 11.973 & 0.620 & -3.2653 & 0.007 &          \\
    V85  &  19:40:49.34  &  40:14:40.06 & 12.270 & 0.544 & 8.5246  & 0.011 &          \\
   V191  &  19:41:42.65  &  40:11:40.42 & 12.819 & 0.288 & 1.6159  & 0.010 &          \\
   V266  &  19:41:11.72  &  40:09:47.48 & 13.028 & 0.413 & 1.0815  & 0.009 &          \\
   V418  &  19:41:45.45  &  40:08:51.03 & 13.366 & 0.634 & 3.5471  & 0.011 &          \\
   V512  &  19:41:08.91  &  40:09:33.36 & 13.530 & 0.646 & 1.3008  & 0.007 &          \\
   V546  &  19:41:35.59  &  40:06:45.78 & 13.616 & 0.681 & 1.9882  & 0.010 &          \\
   V628  &  19:41:17.01  &  40:10:35.03 & 13.764 & 0.686 & 1.1252  & 0.008 & H09 RV member (9003) \\
   V671  &  19:41:16.34  &  40:06:17.16 & 13.825 & 0.633 & 1.0628  & 0.009 &            \\
   V646  &  19:41:29.00  &  40:13:15.79 & 13.828 & 0.572 & 1.4180  & 0.009 & H09 RV member (5006) \\
   V784  &  19:41:03.30  &  40:10:53.07 & 14.040 & 0.368 & 1.6150  & 0.008 & KS88 (V2), T10 (007006)   \\
  V1253  &  19:41:41.75  &  40:15:24.21 & 14.554 & 0.621 & 4.1264  & 0.010 & pulsating $\delta$ Scuti? \\
  V1504  &  19:41:35.48  &  40:16:19.98 & 14.717 & 0.667 & 1.1337  & 0.012 &                           \\
V229980  &  19:41:23.41  &  40:15:17.08 & 14.492 & 0.524 & 1.2180  & 0.009 &                           \\
  V2503  &  19:40:54.91  &  40:01:08.16 & 15.270 & 0.600 & 1.2314  & 0.011 & H09 RV member (32023)     \\  
\enddata
\tablecomments{\footnotesize The numbers in $Note$ represent the IDs of BSS candidates in each study}
\end{deluxetable*}

\section{Summary and Conclusion}

We present $VI$ mosaic CCD photometry of a 1$^\circ$$\times$1$^\circ$ field centered on the old open cluster NGC 6819. Our wide field imaging covers almost the entire dynamical extent of the cluster, and the resultant photometry reaches down to $V$ $\approx$ 21.0 which is $\sim$ 6 mag below the MS turnoff. By analyzing these data, we derived the following results. 

1. Applying the adopted reddening, $E(B-V)=$0.14 and metallicity, $[Fe/H]=+$0.09 dex from B01, we obtained a distance modulus, $(m-M)_{v} =$ 12.36 and an age of 2.6 Gyr for NGC 6819 by fitting theoretical isochrones derived from the Dartmouth Stellar Evolution Database. A change in $E(B-V)$ of $\pm$0.04 mag yields an error in the derived distance modulus of $\pm$0.05 mag and an error in the derived age of $\pm$ 0.5 Gyr with a fixed [Fe/H] value of 0.09 dex. A change in [Fe/H] of $\pm$0.1 dex results in a change in the derived age of $\pm$0.3 Gyr and distance modulus of $\pm$0.08 mag. The direct CMD comparison of our photometry to that of M67 (Montgomery et al. 1993) shows good correspondence and supports our estimate of the distance modulus.  

2. Our radial density profile reveals that the stellar density becomes indistinguishable from the field star density at a radius of $\sim$ 13$'$. This newly determined radius marks the largest cluster extent yet as compared to the measurements from previous studies (i.e., K01 measured a radial extent of $\sim$9.5$'$).  

3. We calculated a tidal mass of the cluster, $M_{tid} =$ 3542.4 $^{+536.8}_{-451.6}$ $M_\odot$, using a modified version of the point-mass Galactic model. By averaging our new estimate of the tidal mass of the cluster with the photometric masses($M_{phot,K01}=$ 2600 $M_\odot$; $M_{phot,this}=$ 2010 $M_\odot$), we obtained the total mass of $<M_c>$ $\sim$ 2700 $M_\odot$ for NGC 6819. 

4. The flattened present-day LF of NGC 6819 indicates that the cluster is dynamically evolved has lost many of its low-mass stars through evaporation. We derived the present-day MF of NGC 6819 using the stellar mass-luminosity relation defined by the theoretical evolutionary tracks of Dotter et al. (2007, 2008). The slope of the global MF ($\Gamma=-$0.65) is shallower than the Salpeter IMF. The radial MFs tends to be more flattened in the central region and appear to be steeper at larger radii, indicating the presence of mass segregation in NGC 6819. The estimated half-mass relaxation time for the cluster ($t_{rh} \sim$ 200 Myr) further supports the assertion that the observed mass segregation in NGC 6819 is a result of its dynamical evolution.

5. Our variability study of the BSSs in NGC 6819 reveals that the majority of the BSS candidates appear to be non-variable star, while some exhibit moderate variability. We also detect a probable $\delta$ Scuti type pulsating BSS using our template light curve fitting routine, RRFIT. Its pulsational properties appear to be consistent with those of the short-period pulsating BSSs found in M67 (Gilland \& Brown 1992).

\acknowledgments

We gratefully acknowledge support for this work from NSF grants AST 98-19768 and AST 0606703.


\begin{thebibliography}{}
\bibitem[]{} Auner, G. 1974, A\&AS, 13, 143
\bibitem[]{} Basu, S., Grundahl, F., Stello, E., et al. 2011, ApJL, 729, L10
\bibitem[]{} Bessell, M. S. 1990, PASP, 102, 1181
\bibitem[]{} Bevington, P. R., and Robinson, D. K. 1992 {\it Data Reduction and Error Analysis For The Physical Sciences, 2nd Edition}, (MacGraw-Hill, Inc.) 
\bibitem[]{} Bragaglia, A., Carretta, E., Gratton, R. G., et al. 2001, AJ, 121, 327 (B01)
\bibitem[]{} Bragaglia A.,\& Tosi M. 2006, AJ, 131, 1544 
\bibitem[]{} Bressen, A., Marigo, P., Giradi, L., et al. 2012, arXiv:1208.4498v1
\bibitem[]{} Burkhead, M. S. 1971, AJ, 79, 251 (B12)
\bibitem[]{} Canterna, R., Geisler, D., Harris, H. C., Olszewski, E., Schommer, R. 1986, AJ, 92, 79 
\bibitem[]{} Carretta et al., 2001
\bibitem[]{} Charbonneau, P. 1995, ApJS, 101, 309
\bibitem[]{} Deliyannis, C. P., 2003, AAS, 203,1420.
\bibitem[]{} Demarque, P., Woo, J.-H., Kim, Y.-C., \& Yi, S. K. 2004, ApJS, 155, 667
\bibitem[]{} Dias W. S., Alessi B. S., Moitinho A. and Lépine J. R. D., 2012yCat.102022
\bibitem[]{} Dotter, A., Chaboyer, B., Ferguson, J. W., et al. 2007, ApJ, 666, 403
\bibitem[]{} Dotter, A., Chaboyer, B., Jevremović, D., et al. 2008, ApJS, 178, 89
\bibitem[]{} Friel, E. E. 1995, ARAA, 33, 381 (D08)
\bibitem[]{} Gilland, R. L., \& Brown, T. M. PASP, 104, 582
\bibitem[]{} Glebbeek, E., Pols, E.,\& Hurley, J. R. 2008, A\&A, 488, 1007
\bibitem[]{} Hauschildt, P. H., Allard, F., \& Baron, E. 1999a, ApJ, 512, 377
\bibitem[]{} Hauschildt, P. H., Allard, F., Ferguson, J., et al. 1999b, ApJ, 525, 871
\bibitem[]{} Hole, K. T., Geller, A. M., Mathieu, R. D., et al. 2009, AJ, 138, 159 (H09)
\bibitem[]{} Innanen, K.A., Harris, W.E. \& Webbink, R.f. 1983 ApJ, 88, 338
\bibitem[]{} Kalirai, J. S., Richer, H. B., Fahlman, G. G., et al. 2001a, AJ, 122, 257
\bibitem[]{} Kalirai, J. S., Richer, H. B., Fahlman, G. G., et al. 2001b, AJ, 122, 266 (K01)
\bibitem[]{} Kaluzny, J., \& Shara, M. M. 1988, AJ, 95, 785 (KS88)
\bibitem[]{} Kang, Y-W., \& Ann, H. B. 2002, JKAS, 35, 87
\bibitem[]{} Keenan, D.W. 1981, A\&A 95,340
\bibitem[]{} Kinemuchi, K., Sarajedini, A., Geisler, D. et al. 2010, IAUS, 266, 429
\bibitem[]{} King, I. 1962, AJ, 67, 471
\bibitem[]{} Lindoff, U. 1972, A\&AS, 7, 497
\bibitem[]{} Mackey, A. D., \& Gilmore, G. F. 2003, MNRAS, 338, 85
\bibitem[]{} Mateo, M. 1993, ASP Conf. Ser. 54, Blue Stragglers, ed. R. A. Saffer (Baltimore, MD: ASP), 74
\bibitem[]{} Mathieu, R. E. 2000ASPC, 198, 517
\bibitem[]{} Montgomery, K. A., Marschall, L., \& Janes, K. A. 1993, AJ, 106, 181
\bibitem[]{} Pamyatnkh, A. A. 2000, ASP Conf. Ser. 210, Delta Scuti and Related Stars, ed. M. Breger \& M. Montgomery (San Francisco CA: ASP), 215 
\bibitem[]{} Rosvick, J. M., \& Vandenberg, D. A., 1998, AJ, 115, 1516 (RV98)
\bibitem[]{} Sarajedini, A., von Hippel, T., Kozhurina-Platais, V., \& Demarque, P. 1999, AJ, 118, 2894
\bibitem[]{} Sarrazine, A. R., Deliyannis, C. P., Sarajedini, A., and Platais, I. 2003, AAS, 203, 1420
\bibitem[]{} Spitzer, L. Jr., Hart, M. H. 1971, ApJ, 166, 483
\bibitem[]{} Spitzer, L. Jr. 1958, ApJ, 127, 17
\bibitem[]{} Spitzer, L. Jr. 1987, {\it Dynamical Evolution of Globular Clusters}, ed. J. P. Osteriker (Princeton, NJ) 
\bibitem[]{} Stello, D., Meibom, S., Gilliland. R. L., et al. 2010, ApJ, 713, L182
\bibitem[]{} Stetson, P. B.,\& Harris, W. E. 1988, AJ, 96, 909
\bibitem[]{} Street, R. A., Horne, K., Lister, T. A., et al. 2002, MNRAS, 330, 737
\bibitem[]{} Street, R. A., Horne, K., Lister, T. A., et al. 2005, MNRAS, 358, 759
\bibitem[]{} Talamentes, A., Sandquist, E., Clem, J. L., et al. 2011, AJ, 140, 1268 (T10)
\bibitem[]{} Welch, E. L., \& Stetson, P. B., 1993, AJ, 105, 1813
\bibitem[]{} Wu, Z-Y., Zhou, X., Ma, J. \& Du, C-H. 2009, MNRAS, 399, 2146
\bibitem[]{} Yang, S-C., \& Sarajedini, A. 2012, MNRAS, 419, 1362

\end{thebibliography}
\end{document}